\documentstyle[aps,epsfig,isolatin1,twocolumn,prl]{revtex}

\newcommand{\chapter}{\section}

\def\comment#1{}

\def\cm#1{}

\everymath={\displaystyle}

\newcommand{\N}{N_c}
\newcommand{\dslash}{\partial\!\!\!/}

\newcommand{\gd}[1]{\gamma_{#1}}  
\newcommand{\ld}[1]{\lambda_{#1}}  
\newcommand{\p}{\partial}          
\newcommand{\f}[2]{\frac{#1}{#2}}
\newcommand{\tr}{{\rm tr}}
\newcommand{\be}{\begin{equation}}
\newcommand{\ee}{\end{equation}}
\newcommand{\beqn}{\begin{eqnarray}}
\newcommand{\eeqn}{\end{eqnarray}}
\newcommand{\Tr}{{\rm Tr}}

\newcommand{\s}{\sigma}

\newcommand{\SigM}{{\rho_0}}

\newcommand{\calL}{\mbox{${\cal L}$}}

\newcommand{\psibar}{\bar{\psi}}

\newcommand{\betacrit}{\beta^{\rm cr}}

\renewcommand{\xi}{M}

\title{No Massless Pions
 in Nambu-Jona-Lasinio Model due to Chiral Fluctuations}

\author{H. Kleinert \and B. Van den Bossche\thanks{On leave from absence of
Physique Nucl\'eaire Th\'eorique, B5, Universit\'e de Li\`ege Sart-Tilman, 4000
Li\`ege, Belgium}}

\address{Institut f\"ur Theoretische Physik, Arnimallee 14 D-14195 Berlin,
 Germany}

\begin{document}

\maketitle

\begin{abstract}
In contrast to common belief, the chirally symmetric Nambu--Jona-Lasinio model
does not  contain massless pions, due to strong chiral fluctuations.
Although quarks acquire spontaneously a nonzero constituent mass $M$,
pions have a nonzero mass equal to the mass of $ \s $-mesons,
both being of the order of $M$. This result is found in several
cutoff schemes. Our derivation is nonperturbative, but involves a simple approximation (London limit)
which should, however, receive only quantitative,
no qualitative corrections.
\end{abstract}

\chapter{Introduction}

The chirally symmetric Nambu--Jona-Lasinio model \cite{NJLM}
 was invented to explain the small mass of the pions
as a result of a spontaneous
 breakdown of
chiral symmetry
in hadron physics. The
first realistic formulation
of the model which included flavored quarks,
possessed chiral symmetry $SU(3) \times SU(3)$,
and a spectrum of $\s, \pi, \rho, A_1$ mesons and their $SU(3)$
partners, was formulated and investigated in 1976 by one of the
 authors~\cite{hadroniz}, and has been followed by many
 papers in nuclear physics in the past twenty-three years
\cite{vdb}.

In two important respects, however, the model was unsatisfactory. First,
it was not renormalizable in four dimensions, but required
a  momentum space cutoff  $ \Lambda $ to
 produce finite results. Moreover, to obtain
physical quantities of the correct size, the cutoff
had  to be rather small, below one~GeV, thus limiting the
applicability of the model to very low energies. Second,  the model could not
account for quark confinement.

The first problem, the nonrenormalizability,
was removed in~\cite{hadroniz} by
replacing the four-fermion interaction by the exchange
of a massive vector meson $V_\mu$. The different attractive meson channels were
obtained by a
Fierz transformation of the effective four-fermion
vector-vector  interaction. The mass
of $V_\mu$ took over the role of the cutoff.
Although this led to a bona-fide field theory,
the range of applicability
was still limited by the second problem, the lack
 quark confinement.

The purpose of this note is to point out a
 much more severe
problem with the model: If chiral fluctuations are  properly taken into
account,
the spontaneous symmetry breakdown disappears, and
the pions acquire a nonzero mass equal to that of the $\s$-mesons.
The argument is nonperturbative, and this is the reason why it has been
overlooked until now.
The chiral properties of the Nambu--Jona-Lasinio model
which have been derived and studied in the existing literature
turn out to exist
only if the
quarks would exist with $N_c>3$ identical replica, instead of the
three colored quarks existing in nature.

The nonperturbative arguments used
in this paper are analogous to those applied before
 \cite{2phts}
 in a discussion of the
Gross-Neveu model~\cite{GNM} in $2+\varepsilon$ dimensions,
where it was shown that this model has two
phase transitions, one where quarks become massive, and another one
where chiral symmetry
is spontaneously broken.

\chapter{Nambu--Jona-Lasinio model}

The model contains $N_f$  massless quark fields $\psi(x)$, each with $N_c$ colors.
Since the fluctuation effects
to be discussed will be caused by the  massless modes, we may
restrict ourselves to   up and down quarks ($N_f=2$).
The Lagrangian of the model is given by \cite{vdb}
\be
\calL=\psibar
i\dslash
\psi+\f{g_0}{2N_c}\left[
\left(
\psibar\psi
\right)^2+\left(
\psibar\ld{a}i\gd{5}\psi
\right)^2
\right].
\label{NJLModel}
\ee
The three $2\times2$-dimensional
matrices $ \ld{a}/2$, generate the fundamental representation
of flavor $SU(2)$, and are normalized by
$\tr (\ld{a}\ld{b})=2\delta_{ab}$.

A Hubbard-Stratonovich transformation makes the model
equivalent to a
theory of  collective scalar and pseudoscalar fields
$ \s$ and $\pi_a$:
\be
\calL=\psibar\left(
i\dslash-\s-i\gd{5}\ld{a}\pi_a
\right)
\psi-\f{\N}{2g_0}\left(
\s^2+\pi_a^2
\right).
\label{hsnjl}
\ee
After integrating out the quark fields, the
limit  $N_c\rightarrow \infty$
leads to an effective action for the ground state
\be
 \Gamma (\rho)=- \Omega [\Delta v( \rho )+v_0]
\label{EffectivePotential}
\ee
where $\Omega$ is the spacetime volume, $v_0$ is the divergent energy density
of the symmetric state,
and $\Delta v(\rho)$
is the condensation energy
\beqn
&&\Delta v(\rho)=\f{N_c}{2}\Bigg\{
\f{1}{g_0}\rho^2-\f{2}{(2\pi)^2}\Bigg[
\f{\rho^2\Lambda^2}{2}
+\f{\Lambda^4}{2}\ln\left(
1+\f{\rho^2}{\Lambda^2}
\right)\nonumber\\
&&~~~~~~~~~\!\mbox{}-\f{\rho^4}{2}\ln\left(
1+\f{\Lambda^2}{\rho^2}
\right)
\Bigg]
\Bigg\}
\label{lambdapotential}
\eeqn
at a constant
$  \sigma ^2+\pi_a^2\equiv  \rho ^2$.
As in the original treatment \cite{NJLM}, the momentum integral is
 regularized by a cutoff
$\Lambda$ in euclidean momentum space.
The condensation energy is extremal
at
$ \rho = \SigM$, which solves the {\em gap equation\/}
\beqn
 \f{1}{g_0}&=&\f{2}{(2\pi)^2}
\left[
\Lambda^2- \SigM^2\ln\left(
1+\f{\Lambda^2}{\SigM^2}
\right)
\right].
\label{lambdagap}
\eeqn
The quantity $\SigM$ plays the role of an {\em order parameter\/}
for the condensed state.
It also
determines the  constituent
mass $M_0$ of the quark fields in the limit $\N\rightarrow \infty$.

\section{Chiral fluctuations}

Since the physical number of quarks
$\N$ is finite, the fields perform fluctuations
 around the extremal field value, which we may assumed to point in
 the $ \sigma $-direction: $( \sigma ,\pi_a)=( \rho _0,0)$.
As long as $\N$ can be considered as a large number,
the deviations
$(\s',\pi'_a)\equiv(\s- \SigM,\pi_a)$
are small, and the action can be expanded in powers
of $(  \s',\pi'_a)$.
The quadratic terms in this expansion
define the propagators of the collective
fields $( \s',\pi'_a)$,
whereas the higher expansion terms
define the interactions.
In momentum space,
the quadratic terms are
\begin{eqnarray}
{\cal A}_0[\s',\pi'] \!=\! \f{1}{2}\!\int\!\!d^4q\!\left[
\left(\!\!
\begin{array}{c}
\pi'_a(q)\\
\s'(q)
\end{array}
\!\!\right)^T\!\!
\!\!\left(\!\!
\begin{array}{cc}
G_{\pi}^{-1}(q)&\!\!\!\!\!\!\!\!0\\
0&\!\!\!\!\!\!\!\!G_{\s}^{-1}(q)
\end{array}
\!\!\right)\!\!
\left(\!\!
\begin{array}{c}
\pi'_a(-q)\\
\s'(-q)
\end{array}
\!\!\right)
\right]\!,\nonumber
\label{@effa}
\label{@}\end{eqnarray}
with the inverse bosonic propagators $G_{\s,\pi}^{-1}(q)$
\begin{displaymath}
\N\bigg\{ 2\times2^{D/2}\!\!\!\int\!\!
\f{d^4p_E}{(2\pi)^4}\f{(p_E^2+p_Eq_E\mp\SigM^2)}
{(p_E^2\!+\!\SigM^2)[(p_E+q_E)^2\!+\!\SigM^2]}
\!-\!\f{1}{g_0}\bigg\}.
\label{@gre1}
\end{displaymath}
With the help of the
gap equation (\ref{lambdagap}), we eliminate the term $1/g_0$
and obtain in $D=4$ dimensions the
inverse propagators
\begin{eqnarray}
\!\!\!\!\!\!\!\!\!\!\!G_{\pi,\sigma}^{-1}(q_E^2)&=&
\N\bigg\{ 8\!\!\int_{0}^1 dy\int_0^{ \Lambda ^2}\!\!
\f{dp_E^2\,p_E^2}{16\pi^2}
\nonumber \\
&\times&
\f{q_E^2(1-y)+(0,2 \rho _0^2)}
{[p_E^2\!+q_E^2y(1-y)+\SigM^2]^2}
\bigg\}.
\label{SigPropStiffLambdax}
\label{@}\end{eqnarray}
For small
momenta, these behave like
\beqn
G_{\pi}^{-1}\approx
Z(\SigM)q_E^2,~~~~
  G_{ \sigma }^{-1}\approx
Z(\SigM)(q_E^2+4\SigM^2),
\label{SigPropStiffLambda}
\eeqn
where $Z( \rho _0)$ is the
wave function renormalization constant
\beqn
Z( \rho _0)=
\f{\N}{(2\pi)^2}
\left[
\ln\left(
1\!+\!\f{\Lambda^2}{\SigM^2}
\right)
\!-\!\f{\Lambda^2}{\Lambda^2\!+\!\SigM^2}\right].
\label{SigPropStiffLambda2}
\eeqn

As a consequence of the spontaneous symmetry breakdown,
the fluctuations of the pseudoscalar fields are massless
{\em Goldstone bosons\/}.
These fields appear in the $x$-space version of the above
quadratic action
in a pure gradient form, thus performing violent fluctuations.
The fluctuations possess a large entropy,
leading to a destruction of the ordered state
and a restoration
of the chiral symmetry, unless $N_c$ is unphysically large.

For simplicity, we shall ignore the massive size fluctuations of the field
$\rho$,
as in  the {\em London limit\/} of superconductivity (also called {\em hydrodynamic limit\/}).
We introduce a unit vector field
$n_i\equiv (n_0,n_a)\equiv(\s,\pi_a)/ \rho $
in the mesonic field space,
whose long-wavelength fluctuations have, for symmetry reasons,
the gradient action
\be
 {\cal A}_0[n_i]= \f{\beta( \rho ) \rho ^2}{2}\int d^4x
[\partial n_i(x)]^2.
\label{@prop}
\ee
The prefactor $\beta( \rho )$ in (\ref{@prop}) is called the {\em stiffness\/}
of the directional
fluctuations \cite{2phts,bkt,cmab,GFCM}.
The first of Eqs.~(\ref{SigPropStiffLambda}) shows that the stiffness is
\be
\beta( \rho )= Z(\rho),
\label{StiffModel}
\ee
to be evaluated at the extremum $  \rho =\rho _0=M$, in the London limit.

For shorter wavelengths,
the action
(\ref{@prop}) becomes
\begin{equation}
{\cal A}_1[n_i]=\frac{\rho ^2}{2}\int d^4x
\,n_i(x)G^{-1}_\pi(-\partial^2 )n_i(x),
\label{@propp}
\end{equation}
with $G^{-1}_\pi(-\partial^2 )$ from Eq.~(\ref{SigPropStiffLambdax}).

We  now demonstrate
that the stiffness (\ref{StiffModel})
in the Nambu--Jona-Lasinio
model for three colored quarks
is far too small to let the spontaneous symmetry
breakdown survive the strong
fluctuations of
the directional field $n_i(x)$.
We do this first
in the long-wave approximation
(\ref{@prop}),  where the discussion
is simplest.
Rewriting
(\ref{@prop}) with the help of a Lagrange multiplier field as
\be
\f{\beta( \rho ) \rho ^2}{2} \int d^4x
\left\{ [\partial n_i(x)]^2+ \lambda(x) \left[ n_i^2(x)-1\right] \right\},
\ee
we integrate out the $n_i(x)$-fields, and find
\be
\tilde{\cal A}_0[ \lambda ]=-\beta( \rho )\rho^2\int d^4x
 \f{\lambda(x)}{2}+\f{N_n}{2}\Tr\ln\left[
-\p^2+\lambda(x)
\right],
\label{newaction}
\ee
where $N_n$ is the number of components of $n_i(x)$, and $\Tr$ denotes the
functional  trace. In going from
${\cal A}_0[n_i]$ to
$\tilde{\cal A}_0[ \lambda ]$, we have summed up infinitely many
diagrams of the ordinary perturbation
theory of
$ \sigma $- and
$\pi_a$-fields, whose expansion parameter is $1/N_c$.

For large $N_n$,
the fluctuations are suppressed. In the ground state, the
field $\lambda(x)$ becomes a constant, satisfying a {\em second gap equation\/}
\be
\beta( \rho )=\frac{N_n}{ \rho ^2}\int \f{d^4k}{(2\pi)^4}\f{1}{k^2+\lambda} .
\label{@secge}\ee
There exists a phase transition at a  critical stiffness
\be
\betacrit=\frac{N_n}{ \rho ^2}\int \f{d^4k}{(2\pi)^4}\f{1}{k^2}.
\label{CriticalStiff}
\ee
For $\beta( \rho )<\betacrit$, chiral fluctuations are so violent that the system
goes into a disordered state with $ \lambda \neq 0$,
where the $N_n$ fields $n_i(x)$ have a nonzero square mass
$\lambda$,
which plays the role of an order parameter in the directional
phase transition.
Since the fields $n_i(x)$ are the normalized $ \sigma $- and $ \pi_a $-fields, $ \lambda $ is a
nonzero common square mass of
the associated particles.
Thus chiral fluctuations
have restored the chiral symmetry which was broken
in the initial large-$N_c$ approximation.

Note that an important consequence of the initial
large-$N_c$ approximation persists:
the quarks are still massive.
This
does not
contradict the Goldstone theorem.
A
nonzero quark mass is perfectly compatible
with chiral symmetry
since
an arbitrary  chiral rotation of the mass term
in the Dirac equation,
performed into any
  pion direction, produces
a term
$m(\cos \chi +i \gamma _5 \sin \chi)$,
and thus describes quarks
of the same mass $ M = \sqrt{ \sigma ^2+ \pi _a^2} $.
Physically, the mass term is a consequence of the {\em formation\/} of
the pairs which for small $N_c$ are strongly bound.
The phase transition taking place at $ \beta( \rho ) = \beta _c$,
on the other hand, describes the Bose condensation of these
pairs, which is a completely different process
for small $N_c$. The separate occurrence of the two transition
follows from
a simple fluctuation criterion \cite{GC}.

In the model, the number $N_n$ is equal to four, such that the large-$N_n$
approximation in the calculation of $ \beta ^{\rm cr}$ may be questioned.
However, Monte-Carlo studies  have shown
that  $N_n=4$
is large enough to ensure the existence of the transition, and that the
critical stiffness obtained from~(\ref{CriticalStiff}) is correct to within
2\%~\cite{montecarl,neuhaus}.

\begin{figure}[h]
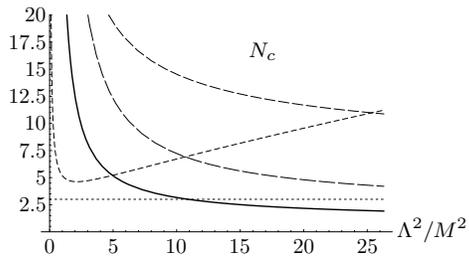

~~~~~\input plotnc.tps
~~\\[-1cm]

\caption[]{Lowest critical number of colors $N_c$
for which there exists a symmetry-broken phase with zero-mass pions
in the Nambu-Jona-Lasinio model, as a function of $ \Lambda ^2/M^2$,
where $ \Lambda $ is the cutoff
for the quark fields with
their constituent mass
 $M$.
The solid line is from the exact propagator with a cutoff
for the pion field $ \Lambda _\pi^2=4M^2$ (approximation 2),
the long-dashed line for $ \Lambda _\pi^2=8M^2$ (approximation 3).
The shorter-dashed curve
comes from an expansion of the denominator of the
exact propagator up to $q^4$ and an infinite cutoff $ \Lambda _\pi$
(approximation 4).
The short-dashed curve
would result from a pure $1/Z( \rho _0) q^2$ approximation
of the pion propagator with a cutoff
$ \Lambda _\pi= \Lambda $
(approximation 1).
Except for an excessively large cutoff $ \Lambda^2\approx11M^2$
in the exact propagator,
which is incompatible with the experimental value of
 $f_\pi\approx 0.093$, the symmetry will always be restered for $N_c=3$ (dotted line).}
\label{fig1n}
\end{figure}

For a first crude estimate of the critical number of colors $N_c^{\rm cr}$
where symmetry restoration
takes place, we cut the divergent integral
(\ref{CriticalStiff}) off at the same $\Lambda$
as the fermion loop integrals. This will be referred to as approximation 1.
Then we obtain,
for $N_n=4$,
\be
\betacrit=4\f{\Lambda^2}{16\pi^2},
\label{StiffNLSigma}
\ee
from which we find, by
comparison with
(\ref{StiffModel}),
\be
N_c^{\rm cr}\!=\!\left(\!\f{\Lambda}{\SigM}\!\right)^2
\!\left\{
\ln\!\left[
1\!+\!\left(
\f{\Lambda}{\SigM}
\right)^2
\right]\!-\!\f{\left(
\Lambda/\SigM
\right)^2}{1\!+\!\left(
\Lambda/\SigM
\right)^2}
\right\}^{-1}.
\label{StiffCond}
\ee
The model
only possesses a
phase in which pions  are Goldstone bosons if the number of colors
exceeds  $N_c^{\rm cr}$.
The critical number (\ref{StiffCond}) is plotted in Fig.~\ref{fig1n} as a
short-dashed curve. We see that $N_c$ would have to exceed the unphysical
number
of
colors 5 to have massless pions.

Let us refine this crude estimate
by going to
approximation 2.
Here we first replace $1/k^2$ in
Eq.~(\ref{CriticalStiff}) by the full pion propagator
$G_\pi(k^2)/ \rho_0 ^2$
associated with the action (\ref{@propp}).
Second, we choose a more physical cutoff $ \Lambda _\pi$
to make
the integral over pion momenta finite.
The pion fields are composite, and will certainly not
be defined over length scales much shorter
than the inverse binding energy of the pair wave function
which is equal to $2M=2\rho_0$.
Thus we perform the integral in the modified Eq.~(\ref{CriticalStiff})
up to the cutoff $4M^2$. This yields the solid curve in
Fig.~\ref{fig1n}.
The intercept of this curve with the $N_c=3$ -line
shows that there exists a phase with broken symmetry for three colors
only if the cutoff of the quark loop integration is $ \Lambda ^2>11M^2$.
Such a large cutoff, however, is incompatible with the experimental value of the pion decay constant
$f_\pi\approx 0.093$. Within the present model, this
constant  is in the large-$N_c$ limit
$f_\pi/M=Z^{1/2}(M)$. For the typical estimates
of constituent quark masses $M\in(300,400)$ MeV \cite{hadroniz},
\comment{{,GL}}
we find that $ \Lambda ^2/M^2$ should lie
in the range $(3.3,7.3)$, the highest value corresponding to the lowest
possible mass 300 MeV.

If we cut the momentum integral over
$G_\pi(k^2)/ \rho_0 ^2$
off at a larger value $8M^2$,
which we call approximation 3,
we obtain the long-dashed curve in
Fig.~\ref{fig1n}. It cuts the  $N_c=3$ -line
at an even higher unphysical value  $ \Lambda ^2/M^2\approx 60$.

Another natural way of cutting off the integral is
by Taylor-expanding the denominator of
$G_\pi(k^2)/ \rho_0 ^2$ up to $k^6$
and doing the, now convergent,  momentum integral
up to infinity.
This yields
approximation 4, pictured
as a  shorter-dashed curve in Fig.~\ref{fig1n}, again with
no symmetry-broken phase for a physically acceptable range of
cutoffs $ \Lambda$.

 Thus we conclude that the Nambu--Jona-Lasinio model
 cannot properly be used to describe
the chiral symmetry breakdown of quark physics
in quantum chromodynamics.

It is interesting to see that the same conclusion cannot be reached
in the dimensional regularization scheme \cite{krew}.
In that scheme, the integral
in (\ref{CriticalStiff})
determining the critical stiffness
vanishes.
This is one of the typical unphysical features of
dimensional regularization \cite{regular},
which can meaningfully be used
only in renormalizable theories, where all quantities
which diverge with a power of the momentum space cutoff $ \Lambda $
can be
absorbed into unobservable bare quantities of the theory,
 thus being
physically irrelevant. In the present nonrenormalizable
theory, they are not!

Let us briefly sketch the calculation
of the
common nonzero mass $ \lambda $ of $\sigma$ and $\pi_a$-fields
in the phase of restored chiral symmetry.
For this we consider the change of the effective potential caused by chiral
fluctuations. They add
to $\Delta v(\rho)$ in~(\ref{EffectivePotential}) the action~(\ref{newaction})
at a constant $\lambda(x)= \lambda $, but with $- \partial ^2$ replaced by
$ G^{-1}_\pi(-\partial^2 )/Z( \rho )$.
\beqn
\Delta' v(\rho,\lambda)&=&-\f{1}{2}
\lambda Z(\rho)\rho^2\nonumber\\
&+&\f{N_n}{2}
\int_0^\infty\frac{dq_E^2\,q_E^2}{16\pi^2}
\log[G^{-1}(q_E^2)/Z(\rho) + \lambda ]
\label{lambdapotential2}
\eeqn
Extremizing $\Delta v(\rho)+\Delta'
v(\rho,\lambda)$ yields two coupled gap equations replacing~(\ref{lambdagap})
and (\ref{@secge}). They can be written  down
explicitly
 for approximation 1. They read
\beqn
x_0\ln\left(1+x_0^{-1}\right)+\f{y}{2}\f{d}{dx}\left(
x\bar{Z}\right)&=&x\ln\left(1+x^{-1}\right),\label{xeqn}\\
N_cx\bar{Z}&=&1-y\ln\left(1+y^{-1}\right)\label{yeqn},
\eeqn
with the reduced quantities $\bar{Z}=\ln\left(1+x^{-1}\right)-(1+x)^{-1}$ and
 $x\equiv\rho^2/\Lambda^2,
y\equiv\lambda/\Lambda^2$. The coupling constant $g_0$ has been eliminated,
 with the help of~(\ref{lambdagap}), in
 favor of the constituent quark mass $\SigM=M_0$ which characterizes
the model uniquely above $N_c^{\rm cr}$.
  For
$\lambda=0$,
Eq.~(\ref{xeqn}) reduces to~(\ref{lambdagap}).
 Equation~(\ref{yeqn}), on the other hand,  determines
 the common square mass $ \lambda $ of $\sigma$ and $\pi_a$ as a function of $N_c$,
which  begins developing for $N_c<N_c^{\rm cr}$.

\comment{Approximation 4 yields similar equations.
The resulting $ \lambda $-curves are plotted in Fig.~\ref{fig2}.}

\comment{The solutions of~(\ref{xeqn}) and~(\ref{yeqn}) are plotted in
Figs.~\ref{fig1}--\ref{fig3} for three different values of
$\rho_0$ corresponding  to
$\SigM>\rho_*$, $\SigM<\rho_*$, $\SigM=\rho_*$, where $\rho_*\approx \sqrt{0.46} \Lambda$
is the
value of $\SigM$ corresponding to the minimum of $N_c^{\rm cr}$
in Fig.~\ref{fig1}, which
corresponds to a constituent
quark mass above $N_c^{\rm cr}=4.62$ of $ 0.678\Lambda$.
The associated solutions
are
represented by the
short-dashed curves in the figures.
The
medium-dashed  curves correspond to a constituent
quark mass $0.479\Lambda$, and the long-dashed to
$ 1.342\Lambda$.
In Fig.~\ref{fig3}, the three curves depend so weakly
on $ \rho _0$ that they seem to coincide.
To make the $ \rho _0$-dependence visible,
we have plotted an extra  dotted curve
for a very small value $\rho_0= 0.224\Lambda$ (dotted).
}

Note that the inclusion of other flavors does not  prevent
the restoration of chiral symmetry,
since the associated pseudoscalar mesons are too massive to make their
fluctuations relevant to the described phenomenon.

The  nonperturbative phenomenon can, of course,
not be reproduced in the typical $1/N_c$-expansions
of the model to any finite order \cite{lemmer}.

\section{Conclusion}

In a  nonperturbative treatment we have shown that
for three colored quarks,
the Nambu--Jona-Lasinio model does not really
display the spontaneous symmetry breakdown
for whose illustration it was invented.
The violent chiral
fluctuations in the degenerate potential valley
on the surface of a sphere in the four-dimensional
field space of $ \sigma $- and $\pi_a$-mesons
restore chiral symmetry, making $\s$ and $\pi$ equally massive,
with a mass of the order of the constituent quark mass.
It will be interesting to see how our limits on the
critical number of colors will change with
various corrections to our approximation.

More details will be published elsewhere.

\begin{acknowledgments}
We thank A. Blotz, E. Babaev, D. Blaschke, M. Borelli, T. Neuhaus, A. Pelster, and
A. Schakel for
discussions. The work of B. VdB was partially supported by the
Institut  Interuniver\-si\-taire des Sciences Nucl\'eaires de Belgique and by the
Alexander von Humboldt Foundation.
\end{acknowledgments}

\newpage

~~\\~~\\

\vspace{1.8cm}

\comment{
\begin{figure}[h]
\unitlength 1cm
\begin{center}
\begin{picture}(10,8)
\put(0,0){\vbox{\begin{center}\psfig{file=lambda.eps,width=9cm}\end{center}}}
\put(0.5,3.5){\footnotesize $\frac{\lambda}{\rho_0^2}$}
\put(4.5,0.25){\footnotesize $N_c$}
\end{picture}
\end{center}
\caption[]{Common square masses $m_{\s}^2=m_{\pi}^2=\lambda$ as a function of $N_c$.
The three curves start at different critical values $N_c^{\rm cr}$
whose values can be read off Fig.~\ref{fig1}.}
\label{fig2}
\end{figure}
 }

\comment{
\begin{figure}[t]
\unitlength 1cm
\begin{center}
\begin{picture}(10,8)
\put(0,0){\vbox{\begin{center}\psfig{file=beta.eps,width=9cm}\end{center}}}
\put(0.5,3.5){\footnotesize $\frac{\beta}{\betacrit}$}
\put(4.38,0.25){\footnotesize $N_c/N_c^{\rm cr}$}
\end{picture}
\end{center}
\caption{Reduced stiffness as a function of $N_c/N_c^{\rm cr}$.
The three curves below $N_c^{\rm cr}$ cannot be distinguished on this scale.
The dotted
  curve corresponds to an extra low value of $\SigM$ to
  show that  below $N_c^{\rm cr}$ the curves deviate from a
straight line.}
\label{fig3}
\end{figure}}


\begin{thebibliography}{999}

\bibitem{NJLM}
 Y. Nambu and G. Jona Lasinio, Phys.~Rev.~{\bf 122}, 345 (1961);
  {\bf 124}, 246 (1961).


\bibitem{hadroniz}
H. Kleinert, {\em On the Hadronization of Quark Theories\/},
     Lectures presented at the Erice Summer Institute 1976, in
     {\em Understanding the Fundamental Constituents of Matter\/},
     Plenum Press, New York, 1978, A. Zichichi ed., pp.~289-390.
     (www.physik.fu-berlin.de/\~{}kleinert/53).

\bibitem{vdb}
For references, see the reviews:
S. P. Klevansky,
Rev. Mod. Phys. {\bf 64}, 649 (1992).\\
T. Hatsuda and T. Kunihiro,
Phys. Rep. {\bf 247}, 221 (1994).\\
B. Van den Bossche,
{\em A chiral Lagrangian with three flavors, axial and scale anomalies}
(nucl-th/9807010).

\bibitem{2phts}
H. Kleinert and E. Babaev
Phys. Lett. B {\bf 438}, 311 (1998)

\bibitem{GNM}
{D.~Gross} and {A.~Neveu}, Phys.~Rev.~D {\bf 10}, 3235 (1974).
The model had been discussed earlier by
{V.~G.~Vaks} and A.~I.~Larkin, JETP (Sov.~Phys.) {\bf 13}, 979 (1961),
and by
A.~A.~Anselm, JETP (Sov.~Phys.) {\bf 9\/}, 608 (1959).

\cm{\bibitem{QM} H. Kleinert,
     Phys.~Letters   B {\bf 62}, 77 (1976).}

\bibitem{bkt}
V. L. Berezinskii.
 Zh.~Eksp.~Teor.~Fiz., 1970, vol. 59, No~3, p.907 - 920; \\
J. Kosterlitz, D. Thouless.
J.~Phys., 1973, vol. C6, No
7, p.~1181 - 1203.
\bibitem{cmab}
H. Kleinert
 \cm{
{\em Theory of Fluctuating Nonholonomic Fields and Applications:
    Statistical Mechanics of Vortices and Defects and
 New Physical
      Laws in Spaces with Curvature and Torsion\/},}
        in: {\em Proceedings of a NATO Advanced Study Institute on
        Formation and Interactions of Topological Defects\/}
at the University of Cambridge, England
 (cond-mat/9503030) (www.physik.fu-berlin.de/\~{}kleinert/227).



\bibitem{GFCM}
H.~Kleinert, {\em Gauge Fields in Condensed Matter\/},
World Scientific,~1989.
(www.physik.fu-berlin.de/\~{}kleinert/re0.\linebreak html\#b1).



\bibitem{GC}
  H. Kleinert
{\em Criterion for Dominance of Directional versus Size Fluctuations of Order
Field in Restoring Spontaneously Broken Continuous Symmetries\/},
Berlin Preprint 1999,
(cond-mat/9908239)




\bibitem{montecarl}
Simulations of the four-dimensional $0(4)$ model on a simple-cubic lattice
gives $\betacrit\approx 0.6090$ \cite{neuhaus}.
This is to be compared with the approximation

(\ref{StiffNLSigma})
 calculated for a
simple-cubic lattice, where
$ \beta ^{\rm cr}=4\times0.1549\approx0.6196$
which is thus correct to within less than 2\%. The value 0.1549 is taken from
the list
of
lattice
Coulomb potentials at the origin in Table~6.4
of the textbook \cite{GFCM} on p.~178.

\cm{\bibitem{kanaya}
Simulations of the three-dimensional $0(4)$ model on a simple cubic lattice
gives $\betacrit\approx 0.936$ \cite{kanaya2}. The same value is
obtained by high temperature expansions~\cite{butera}.
This is to be compared with the approximation
 (\ref{StiffNLSigma})
 calculated on the
simple-cubic lattice, where
$ \beta ^{\rm cr}=4\times0.2527\approx1.01$
which differs by less than 10\%. The value 0.2527 is taken from the list
of
lattice
Coulomb potentials at the origin in Table~6.4
of the textbook \cite{GFCM} on p.~178.
%
\bibitem{kanaya2}
K. Kanaya, S. Kaya, Phys. Rev. D {\bf 51}, 2404 (1995);\\
 H. G. Ballesteros, L. A. Fernandez, V. Martin-Mayor, A. Munoz Sudupe,
Phys. Lett. B {\bf 387}, 125 (1996).
%
\bibitem{butera}
P. Butera, M. Comi, Phys. Rev. B {\bf 56}, 8212 (1997).}

\bibitem{neuhaus}
A. Hasenfratz, K. Jansen, J. Jersak, H.A. Kastrup, C.B. Lang, H. Leutwyler,
T. Neuhaus, Nucl. Phys. B {\bf 356}, 332 (1991)

\comment{\bibitem{GL}
 J.~Gasser and H.~Leutwyler,
Phys.~Rept.~{\bf 87}, 77 (1982).}


\bibitem{krew}
S. Krewald and K. Nakayama, Ann. Phys. {\bf 216}, 201 (1991).
\bibitem{regular}
As a prominent unphysical feature, consider
the loop integral for the vacuum energy
in three dimensions
$
{\scriptscriptstyle\int}$$ d^3p \ln(1+m^2/p^2)\propto\left\{
\Lambda^3\ln[1+(m/\Lambda)^2]\right.$ $
\left.+2m^2[\Lambda-m\arctan(\Lambda/m)]\right\}
\sim3m^2\Lambda-\pi m^3+{\cal O}(1/ \Lambda ).
$
The dimensionally regulized integral yields only the
finite negative energy $-\pi m^3$, but fails to find
the physical large positive energy
$3m^2 \Lambda $ of the vacuum fluctuations.

\bibitem{lemmer}
V. Dmitra{\v s}inovi{\'c}, H.-J. Schulze, R. Tegen and
R. H. Lemmer,
Ann. Phys. {\bf 238}, 332 (1994);
Phys. Rev. D {\bf 52}, 2855 (1995).
\end{thebibliography}
\end{document}